\documentclass[sigconf]{acmart}

\usepackage{url}
\usepackage[utf8]{inputenc}
\usepackage{amsmath}
\usepackage{amssymb}
\usepackage{amsthm}
\usepackage{stmaryrd}
\usepackage{tikz}
\usepackage{array}
\usepackage{caption}
\usepackage{tabularx}
\usetikzlibrary{automata,positioning}

\copyrightyear{2018} 
\acmYear{2018} 
\setcopyright{acmlicensed}
\acmConference[SAC 2018]{SAC 2018: Symposium on Applied Computing }{April 9--13, 2018}{Pau, France}
\acmPrice{15.00}
\acmDOI{10.1145/3167132.3167261}
\acmISBN{978-1-4503-5191-1/18/04}

\begin{document}

\title{Syntax Error Recovery in Parsing Expression Grammars}

\author{Sérgio Medeiros}
\affiliation{UFRN, Brazil}
\email{sergiomedeiros@ect.ufrn.br}
\author{Fabio Mascarenhas}
\affiliation{UFRJ, Brazil}
\email{fabiom@dcc.ufrj.br}

\begin{abstract}
Parsing Expression Grammars (PEGs) are a formalism used
to describe top-down parsers with backtracking. As PEGs do
not provide a good error recovery mechanism, PEG-based parsers
usually do not recover from syntax errors in the input,
or recover from syntax errors using ad-hoc, implementation-specific
features. The lack of proper error recovery makes PEG parsers 
unsuitable for using with Integrated Development Environments 
(IDEs), which need to build syntactic trees even for incomplete,
syntactically invalid programs.

We propose a conservative extension, based on PEGs with labeled failures,
that adds a syntax error recovery
mechanism for PEGs. This extension associates {\em recovery expressions}
to labels, where a label now not only reports a syntax error but
also uses this recovery expression to reach a synchronization
point in the input and resume parsing.
We give an operational semantics of PEGs with this recovery
mechanism, and use an implementation based on such semantics to
build a robust parser for the Lua language. We evaluate
the effectiveness of this parser, alone and in comparison with
a Lua parser with automatic error recovery generated by ANTLR, a
popular parser generator.
\end{abstract}


\maketitle


\newcommand{\fivespaces}{\;\;\;\;\;}
\newcommand{\tenspaces}{\fivespaces\fivespaces}
\newcommand{\twentyspaces}{\tenspaces\tenspaces}
\newcommand{\mylabel}[1]{\, \mathbf{(#1)}}
\newcommand{\Lp}{\stackrel{\mbox{\tiny{PEG}}}{\leadsto}}
\newcommand{\Lpp}{\stackrel{\mbox{\tiny{PEGF}}}{\leadsto}}
\newcommand{\Lpl}{\stackrel{\mbox{\tiny{PEGL}}}{\leadsto}}
\newcommand{\throw}{\Uparrow}
\newcommand{\Tup}[2]{(#1,\,#2)}
\newcommand{\Trip}[3]{(#1,\,#2,\,#3)}
\newcommand{\Xp}{x^\prime}
\newcommand{\Xpp}{x^{\prime\prime}}
\newcommand{\Yp}{y^\prime}
\newcommand{\dplus}{\ensuremath{\mathbin{+\mkern-5mu+}}}
\newcommand{\Zp}{z^\prime}
\newcommand{\Any}{X}
\newcommand{\Ex}{v?}
\newcommand{\Ey}{w?}
\newcommand{\Suf}{{\tt min}}
\newcommand{\Suff}[2]{\Suf(#1,\,#2)}
\newcommand{\Lfail}{{\tt fail}}
\newcommand{\Llfail}{l^{\Fail}}
\newcommand{\Nil}{{\tt nil}}
\newcommand{\List}{L}
\newcommand{\Mon}[1]{\{#1\}}
\newcommand{\J}{{\tt join}}
\newcommand{\Jj}[2]{\J(#1,\,#2)}
\newcommand{\Jc}{{\tt joinCon3}}
\newcommand{\Jcc}[2]{\Jc(#1,\,#2)}
\newcommand{\Jv}{{\tt joinVar}}
\newcommand{\Jvv}[1]{\Jv(#1)}
\def\drawplusplus#1#2#3{\hbox to 0pt{\hbox to #1{\hfill\vrule height #3 depth
      0pt width #2\hfill\vrule height #3 depth 0pt width #2\hfill
      }}\vbox to #3{\vfill\hrule height #2 depth 0pt width
      #1 \vfill}}
\def\concat{\mathrel{\drawplusplus {12pt}{0.4pt}{8pt}}}

\newcommand{\Peg}[2]{#1[#2]}
\newcommand{\Pgg}[1]{\Peg{G}{#1}}
\newcommand{\Mat}[2]{#1\;\,#2\,}
\newcommand{\Matg}[2]{\Mat{\Pgg{#1}}{#2}}
\newcommand{\Matk}[3]{#1\;\,#2\;\,#3\,}
\newcommand{\Matgk}[3]{\Matk{\Pgg{#1}}{#3}{#2}}
\newcommand{\Rec}{\slash\!\slash}

\def\verylongrightarrow{\hbox to 35pt{\rightarrowfill}}

\newcommand{\Pow}{\, \widehat{} \,\,}
\newcommand{\Matchf}[4]{\mathrm{match} \;\, #1\;#2\;#3\;#4}
\newcommand{\Matchl}[4]{\mathrm{match}_L \;\, #1\;#2\;#3\;#4}
\newcommand{\Justc}[2]{(\mbox{#1},\,#2)}
\newcommand{\Striple}[3]{(#1,\,#2,\,#3)}
\newcommand{\Sstate}[3]{\langle#1,\,#2,\,#3\rangle}
\newcommand{\Sstatel}[4]{\langle#1,\,#2,\,#3,\,#4\rangle}
\newcommand{\Sstatec}[5]{\langle#1,\,#2,\,#3,\,#4,\,#5\rangle}
\newcommand{\Sfail}[1]{\mbox{\bf Fail}{\langle#1\rangle}}
\newcommand{\Sstep}[4]{#1 & \stackrel{#2}{\verylongrightarrow} & #3 #4\\}
\newcommand{\Sstepf}[4]{#1 & \stackrel{#2}{\verylongrightarrow} & #3 #4}
\newcommand{\Ssteppp}[3]{#1 & \stackrel{\parbox{10pt}{\addvspace{5pt}}}{\verylongrightarrow} & #2 #3\\}
\newcommand{\Sstepppf}[3]{#1 & \stackrel{\parbox{10pt}{\addvspace{5pt}}}{\verylongrightarrow} & #2 #3}
\newcommand{\Sstepp}[4]{#1 \xrightarrow{#2} #3\\}
\newcommand{\Ssteppf}[4]{#1 \xrightarrow{#2} #3}
\newcommand{\Ia}[2]{\mbox{{\scriptsize {\tt #1} $#2$}}}
\newcommand{\Iaa}[3]{\mbox{{\scriptsize {\tt #1} $#2$ $#3$}}}
\newcommand{\In}[2]{\mbox{{{\tt #1} $#2$}}}
\newcommand{\Inl}[3]{\mbox{{{\tt #1} $#2$ $#3$}}}
\newcommand{\Fail}{\mbox{\bf Fail}}
\newcommand{\Hd}[1]{\mbox{hd}(#1)}
\newcommand{\Sub}{{s}}
\newcommand{\Subb}{{s_1}}
\newcommand{\Subi}{\Sub [i]}
\newcommand{\Subbi}{\Subb [i]}
\newcommand{\MathN}[1]{\mbox{\emph{#1}}}
\newcommand{\Nat}{\mbox{\bf N}}
\newcommand{\myarrow}{\xrightarrow{\;\;\;\;\;}}
\newcommand{\myarrowstar}{\xrightarrow{\;\;*\;\;}}
\newcommand{\chmath}[1]{\mbox{`}#1\mbox{'}}

\newcommand{\Sfaill}[2]{\mbox{\bf Fail}{\langle#1,\,#2\rangle}}
\newcommand{\Sfaills}[3]{\mbox{\bf Fail}{\langle#1},\,#2,\,#3\rangle}
\newcommand{\Sfailrec}[4]{\mbox{\bf Fail}{\langle#1},\,#2,\,#3,\,#4\rangle}
\newcommand{\Sthrowks}[5]{\mbox{\bf Throw}{\langle#1},\,#2,\,#3,\,#4,\,#5\rangle}
\newcommand{\Sfailf}[1]{\mbox{\bf Fail}{\langle\Lfail,\,#1\rangle}}
\newcommand{\Sfailfs}[2]{\mbox{\bf Fail}{\langle\Lfail,\,#2,\,#1\rangle}}
\newcommand{\Sfailfrec}[3]{\mbox{\bf Fail}{\langle\Lfail,\,#2,\,#1,\,#3\rangle}}
\newcommand{\Sfailfks}[3]{\mbox{\bf Fail}{\langle\Lfail,\,#2,\,#3,\,#1\rangle}}

\newcommand{\Epsi}{\varepsilon}

\section{Introduction} \label{sec:intro}

Parsing Expression Grammars (PEGs)~\cite{ford2004peg} are a
formalism for describing the syntax of programming languages.
We can view a PEG as a formal description of a top-down parser
for the language it describes. PEGs have a concrete syntax based on
the syntax of {\em regexes}, or extended regular expressions.
Unlike Context-Free Grammars (CFGs),
PEGs avoid ambiguities in the definition of the grammar's
language due to the use of an {\em ordered choice} operator.

More specifically, a PEG can be interpreted as the specification
of a recursive descent parser with restricted (or local) backtracking.
This means that the alternatives of a choice are tried in
order; when the first alternative recognizes an input prefix,
no other alternative of this choice is tried, but when an
alternative fails to recognize an input prefix, the parser
backtracks to try the next alternative.

A naive interpretation of PEGs is problematic when dealing with
inputs with syntactic errors, as a failure during parsing an input
is not necessarily an error, but can be just an indication that the
parser should backtrack and try another alternative. While
PEGs cannot use error handling techniques that are often applied to
predictive top-down parsers, because these techniques assume the parser
reads the input without backtracking~\cite{ford2002packrat,michael2016error},
some techniques for correctly reporting syntactic errors in PEG parsers 
have been proposed, such as tracking the position
of the farthest failure~\cite{ford2002packrat} and
labeled failures~\cite{maidl2013peglabel,maidl2016peglabel}.

While these error reporting techniques improve the quality of
error reporting in PEG parsers, they all assume that the parser
aborts after reporting the first syntax error. While we believe
this is acceptable for a large class of parsers, 
Integrated Development Environments (IDEs) often require parsers
that can recover from syntax errors and build syntax trees
even for syntactically invalid programs, in other to conduct
further analyses necessary for IDE features such as auto-completion.
These parsers should also be fast, as the user of an IDE expects
an almost instantaneous feedback.

Some PEG parser generators already provide some ad-hoc mechanisms
that can be exploited to perform error recovery\footnote{See
threads \url{https://lists.csail.mit.edu/pipermail/peg/2014-May/000612.html}
and \url{https://lists.csail.mit.edu/pipermail/peg/2017-May/000719.html}  of the PEG mailing list,},
but the mechanisms are specific to each implementation, tying the grammar to a specific
implementation. To address this issue, we present a conservative
extension of PEGs, based on labeled failures, that adds a recovery
mechanism to the PEG formalism itself. The mechanism attaches
recovery expressions to labels so that throwing those labels
not only reports syntax errors but also skips the erroneous input
until reaching a synchronization point and resuming parsing.

We give an operational semantics of PEGs with this recovery mechanism
and use an implementation based on such semantics to build a robust
parser for the Lua language. Then we compare the error recovery behavior
of this parser with a Lua parser generated with ANTLR~\cite{parr2011llstar,parr2014antlr},
a popular parsing tool based on a top-down approach.

The remainder of this paper is organized as follows: the next
section (Section~\ref{sec:pegs}) revisits the error handling problem
in PEG parsers and introduces labeled PEGs with our recovery mechanism; 
Section~\ref{sec:strat} discusses error recovery strategies
that PEG-based parsers can implement using our recovery mechanism;
Section~\ref{sec:eval} evaluates our error recovery
approach by comparing a PEG-based parser for the Lua language
with an ANTLR-generated parser;
Section~\ref{sec:rel} discusses related work on error recovery
for top-down parsers with backtracking;
finally, Section~\ref{sec:conc} gives some
concluding remarks.

\section{PEGs with Error Recovery}
\label{sec:pegs}

\begin{figure}[t]
{\small
\centering
\begin{align*}
\it{Prog} & \leftarrow {\tt PUBLIC\;CLASS\;NAME\;LCUR\;PUBLIC\;STATIC\;VOID\;MAIN}\\
& \;\;\;\;\;{\tt LPAR\;STRING\;LBRA\;RBRA\;NAME\;RPAR}\;\it{BlockStmt}\;{\tt RCUR}\\
{\it BlockStmt} & \leftarrow {\tt LCUR}\;{\it (Stmt)*}\;{\tt RCUR}\\
{\it Stmt} & \leftarrow {\it  IfStmt\;/\;WhileStmt\;/\; PrintStmt\;/\;DecStmt\;/\;AssignStmt\;/}\\
& \;\;\;\;\;{\it BlockStmt}\\
{\it IfStmt} & \leftarrow {\tt IF\;LPAR}\;{\it Exp}\;{\tt RPAR}\;{\it Stmt}\;{\it (}{\tt ELSE}\;{\it Stmt\;/\;\varepsilon)}\\
{\it WhileStmt} & \leftarrow {\tt WHILE\;LPAR}\;{\it Exp}\;{\tt RPAR}\;{\it Stmt}\\
{\it DecStmt} & \leftarrow {\tt INT\;NAME}\;({\tt  ASSIGN}\;{\it Exp}\;/\;\varepsilon)\;{\tt SEMI}\\
{\it AssignStmt} & \leftarrow {\tt NAME\;ASSIGN}\;{\it Exp}\;{\tt SEMI}\\
{\it PrintStmt} & \leftarrow {\tt PRINTLN\;LPAR}\;{\it Exp}\;{\tt RPAR\;SEMI}\\
\it{Exp} & \leftarrow \it{RelExp \; ({\tt EQ} \; RelExp)*} \\
\it{RelExp} & \leftarrow \it{AddExp \; ({\tt LT} \; AddExp)*} \\
\it{AddExp} & \leftarrow \it{MulExp \; (({\tt PLUS} \; / \; {\tt MINUS}) \; MulExp)*} \\
\it{MulExp} & \leftarrow \it{AtomExp \; (({\tt TIMES} \; / \; {\tt DIV}) \; AtomExp)*} \\
\it{AtomExp} & \leftarrow \it{{\tt LPAR} \; Exp \; {\tt RPAR} \; / \; {\tt NUMBER} \; / \; {\tt NAME}}
\end{align*}
}
\vspace{-0.7cm}
\caption{A PEG for a tiny subset of Java}
\label{fig:javagrammar}
\end{figure}

In this section, we revisit the problem of error handling
in PEGs, and show how labeled
failures~\cite{maidl2013peglabel,maidl2016peglabel}
combined with the farthest failure heuristic~\cite{ford2002packrat}
can improve the error messages of a PEG-based parser.
Then we show how labeled PEGs can be the basis of
an error recovery mechanism for PEGs, and show an extension
of previous semantics for labeled PEGs that adds
recovery expressions.

\subsection{PEGs and Error Reporting}

A PEG $G$ is a tuple $(V,T,P,p_{S})$ where $V$ is a finite set of
non-terminals, $T$ is a finite set of terminals, $P$ is a total
function from non-terminals to \emph{parsing expressions} and $p_{S}$
is the initial parsing expression.
We describe the function $P$ as a set of rules of the form
$A \leftarrow p$, where $A \in V$ and $p$ is a parsing
expression.
A parsing expression, when applied to an input string, either
fails or consumes a prefix of the input and returns the
remaining suffix. The abstract syntax of parsing expressions is
given as follows, where $a$ is a terminal, $A$ is a non-terminal,
and $p$, $p_1$ and $p_2$ are parsing expressions:
\[\it{
p = \varepsilon \; | \; a \; | \; A \; | \; p_1 p_2 \; | \;
  p_1 / p_2 \; | \; p\!* \; | \; !p
}\]

Intuitively,
$\varepsilon$ successfully matches the empty string, not changing
the input;
$a$ matches and consumes itself or fails otherwise;
$A$ tries to match the expression $P(A)$;
$p_1 p_2$ tries to match $p_1$ followed by $p_2$;
$p_1 / p_2$ tries to match $p_1$;
if $p_1$ fails, then it tries to match $p_2$;
$p*$ repeatedly matches $p$ until $p$ fails, that is, it
consumes as much as it can from the input;
the matching of $!p$ succeeds if the input does not match $p$
and fails when the input matches $p$, not consuming any input in either case;
we call it the negative predicate or the lookahead predicate.

Figure~\ref{fig:javagrammar} shows a PEG for a tiny subset of Java, where
lexical rules (shown in uppercase) have been elided. While simple (this
PEG is equivalent to an LL(1) CFG),
this subset is already rich enough to show the problems of PEG error
reporting; a more complex grammar for a larger language just compounds these problems.

Figure~\ref{fig:javaerror} is an example of Java program
with two syntax errors (a missing semicolon at the end of line
7, and an extra semicolon at the end of line 8).
A predictive top-down parser will detect the first error
when reading the {\tt RCUR} ({\tt \}}) token at the beginning of line 8,
and will know and report to the user that it was expecting a semicolon.

\begin{figure}[t]
{\small
\begin{verbatim}
    1  public class Example {
    2    public static void main(String[] args) {
    3      int n = 5;
    4      int f = 1;
    5      while(0 < n) {
    6        f = f * n;
    7        n = n - 1
    8      };
    9      System.out.println(f);
    10   }
    11 }
\end{verbatim}
}
\caption{A Java program with a syntax error}
\label{fig:javaerror}
\end{figure}

In the case of our PEG, it will still fail when trying to
parse the {\tt SEMI} rule, which should match a {\tt `;'},
while the input has a closing curly bracket,
but as a failure does not guarantee the presence of an error the parser
cannot report this to the user. Failure during parsing of a PEG usually
just means that the PEG should backtrack and try a different
alternative in an ordered choice, or end a repetition. 
For example, three failures will
occur while trying to match the {\it BlockStmt} rule inside {\it Prog} against the {\tt n} at the beginning of line 3, first against {\tt IF} in the {\it IfStmt} rule, then against {\tt WHILE} in the {\it WhileStmt} rule, and finally against {\tt PRINTLN} in the {\it PrintStmt} rule.

After all the failing and backtracking, the PEG in our example
will ultimately fail in the {\tt RCUR} rule of the initial {\it BlockStmt},
after
consuming only the first two statements of the body of {\tt main}.
Failing to match the
{\tt SEMI} in {\it AssignStmt} against the closing curly bracket in the input
will make the PEG backtrack to the beginning of the statement to try
the other alternatives in {\it Stmt}, which also fail. This
marks the end of the repetition inside the {\it BlockStmt} that
is parsing the body of the {\tt while} statement. The whole {\it BlockStmt}
will fail trying to match {\tt RCUR} against the {\tt n}
in the beginning of line 7, this ultimately makes the whole 
{\it WhileStmt} fail, which makes the PEG backtrack to the beginning
of line 5. Now the process repeats with the {\it BlockStmt} that
is parsing the body of {\tt main}.

In the end, the PEG will report that it failed and cannot proceed
at the beginning of line 5, complaining that the {\tt while} in the
input does not match the {\tt RCUR} that it expects, which does not
help the programmer in finding and fixing the actual error.

To circumvent this problem, Ford~\cite{ford2002packrat} suggested
that the furthest position in the input where a failure has occurred
should be used for reporting an error. A similar approach
for top-down parsers with backtracking was also suggested by
Grune and Jacobs~\cite{grune2010ptp}. 

\begin{figure}[t]
{\small
\begin{align*}
\it{Prog} & \leftarrow {\tt PUBLIC\;CLASS\;NAME\;LCUR\;PUBLIC\;STATIC\;VOID\;MAIN} \\
& \;\;\;\;\;\;\;\;{\tt LPAR\;STRING\;LBRA\;RBRA\;NAME\;RPAR}\;\it{BlockStmt}\;{\tt RCUR}\\
{\it BlockStmt} & \leftarrow {\tt LCUR}\;{\it (Stmt)*}\;[{\tt RCUR}]^{\tt rcblk}\\
{\it Stmt} & \leftarrow {\it  IfStmt\;/\;WhileStmt\;/\; PrintStmt\;/\;DecStmt\;/\;AssignStmt\;/}\\
& \;\;\;\;\;\;\;\;{\it BlockStmt}\\
{\it IfStmt} & \leftarrow {\tt IF\;[LPAR]^{\tt lpif}}\;[{\it Exp}]^{\tt condi}\;[{\tt RPAR}]^{\tt rpif}\;[{\it Stmt}]^{\tt then}\\
& \;\;\;\;\;\;\;\;{\it (}{\tt ELSE}\;[{\it Stmt}]^{\tt else}\;/\;\varepsilon)\\
{\it WhileStmt} & \leftarrow {\tt WHILE\;[LPAR]^{\tt lpw}}\;[{\it Exp}]^{\tt condw}\;[{\tt RPAR}]^{\tt rpw}\;[{\it Stmt}]^{\tt body}\\
{\it DecStmt} & \leftarrow {\tt INT}\;[{\tt NAME}]^{\tt ndec}\;({\tt ASSIGN}\;[{\it Exp}]^{\tt edec}\;/\;\varepsilon)\;[{\tt SEMI}]^{\tt semid}\\
{\it AssignStmt} & \leftarrow {\tt NAME\;[ASSIGN]^{\tt assign}}\;[{\it Exp}]^{\tt rval}\;[{\tt SEMI}]^{\tt semia}\\
{\it PrintStmt} & \leftarrow {\tt PRINT\;[LPAR]^{\tt lpp}}\;[{\it Exp}]^{\tt eprint}\;{\tt [RPAR]^{\tt rpp}\;[SEMI]^{\tt semip}}\\
\it{Exp} & \leftarrow \it{RelExp \; ({\tt EQ} \; [RelExp]^{\tt relexp})*} \\
\it{RelExp} & \leftarrow \it{AddExp \; ({\tt LT} \; [AddExp]^{\tt addexp})*} \\
\it{AddExp} & \leftarrow \it{MulExp \; (({\tt PLUS} \; / \; {\tt MINUS}) \; [MulExp]^{\tt mulexp})*} \\
\it{MulExp} & \leftarrow \it{AtomExp \; (({\tt TIMES} \; / \; {\tt DIV}) \; [AtomExp]^{\tt atomexp})*} \\
\it{AtomExp} & \leftarrow \it{{\tt LPAR} \; [Exp]^{\tt parexp} \; [{\tt RPAR}]^{\tt rpe} \; / \; {\tt NUMBER} \; / \; {\tt NAME}}
\end{align*}
}
\vspace{-0.7cm}
\caption{A PEG with labels for a small subset of Java}
\label{fig:javalabels}
\end{figure}

In our previous example, the use of the farthest failure approach
reports an error at the beginning of line 8, the same as a predictive
parser would. We can even use a map of lexical rules to token names to track
expected tokens in the error position, and report that a semilocon
was expected.

If the programmer fixes this error, the parser will then fail repeatedly
at the extra semicolon at line 8, while trying to match the first term
of all the alternatives of {\it Stmt}. This will end the repetition
inside {\it BlockStmt}, and then another failure will happen when trying to match a {\tt RCUR} token against the semicolon, finally aborting the parse.
The parser can use the furthest failure information to report
an error at the exact position of the semicolon, and a list
of expected tokens that includes {\tt IF}, {\tt WHILE}, {\tt NAME}, {\tt LCUR},
{\tt PRINTLN}, and {\tt RCUR}.

The great advantage of using the farthest failure is that the grammar writer does not need to do anything to get a parser with better error reporting,
as the error messages can be generated automatically.
However, although this approach gives us error messages with a
fine approximation of the error location, these messages may not give
a good clue about how to fix the error, and may contain a long list of expected tokens~\cite{maidl2016peglabel}.

We can get more precise error messages at the cost of manually annotating
the PEG with {\em labeled failures}, a conservative extension of the
PEG formalism. A labeled PEG $G$ is a tuple $(V,T,P,L,{\tt fail}, p_{S})$ where
$L$ is a finite set of labels, ${\tt fail} \notin L$ is a failure label,
and the expressions in $P$ have been extended with the
{\em throw} operator, represented by $\throw$.
The parsing expression $\throw^{l}$, where $l \in L$,
generates a failure with label $l$.

A label $l \neq {\tt fail}$ thrown by $\throw$ cannot be caught by an ordered
choice\footnote{This is a simplification of the original
formalization of labeled failures, where a new labeled ordered choice
expression could catch labels; labeled ordered choice is not necessary when
using labels for error reporting.},
so it indicates an actual error during parsing, while {\tt fail}
is caught by a choice and indicates that the parser should backtrack.
The lookahead operator $!$ captures any label and turns it into a success,
while turning a success into a {\tt fail} label.

We can map different labels to different error messages, and then
annotate our PEG with these labels. Figure~\ref{fig:javalabels}
annotates the PEG of Figure~\ref{fig:javagrammar} (except for the
{\it Prog} rule). 
The expression $[p]^{l}$ is syntactic sugar for $(p \; / \; \throw^{l})$. 

The strategy we used to annotate the grammar was the following:
on the right-hand side of a production, we annotate every symbol (terminal
or non-terminal) that should not fail, that is, making the PEG
backtrack on failure of that symbol would
be useless, as the whole parse would either fail or not consume
the whole input in that case. For an LL(1) grammar like the
one in our example, that means all symbols in the right-hand side
of a production except the one in the very beginning of the production.
We apply a similar rule when the right-hand side has a choice or a repetition as a 
subexpression. 

Using this labeled PEG in our program, the first syntax error now
fails directly with a {\tt semia} label, which we can map to
a ``missing semicolon in assignment" message. If the programmer fixes this,
the second error will fail with a {\tt rcblk} label, which
we can map to a ``missing end of block" message.

Compared with the farthest failure approach, one drawback of
labeled failures is the annotation burden. But we can combine
both approaches, and still track the position, and set of
expected lexical rules, of the furthest simple failure.
The parser can fall back on automatically generated error messages
whenever parsing fails without producing a more specific error label.

\subsection{Error Recovery}
\label{subsec:errorrec}

\begin{figure*}[t]
	{\small
		\begin{align*}
		& \textbf{Empty} \;\;\;
		{\frac{}{G[\varepsilon] \; R \; x \Lp \Trip{x}{\Nil}{[]}}} \mylabel{empty.1}
		\fivespaces \textbf{Non-terminal} \;\;\;
		{\frac{G[P(A)] \; R \; xy \Lp \Trip{y}{\Ex}{L}}
			{G[A] \; R \; xy \Lp \Trip{y}{\Ex}{L}}} \mylabel{var.1}
		\;\;\;
		{\frac{G[P(A)] \; R \; x \Lp \Trip{l}{\Ex}{L}}
			{G[A] \; R \; x \Lp \Trip{l}{\Ex}{L}}} \mylabel{var.2}	
		\\ \\
		& \textbf{Terminal} \fivespaces
		{\frac{}{G[a] \; R \; ax \Lp \Trip{x}{\Nil}{[]}}} \mylabel{term.1} 
		\fivespaces
		{\frac{}{G[b] \; R \; ax \Lp \Trip{\Lfail}{ax}{[]}}}, b \ne a \mylabel{term.2}
		\fivespaces
		{\frac{}{G[a] \; R \; \varepsilon \Lp \Trip{\Lfail}{\varepsilon}{[]}}} \mylabel{term.3}
		\\ \\
		& \textbf{Sequence} \;\;
		{\frac{G[p_1] \; R \; xyz \Lp \Trip{yz}{\Ex}{L_1} \fivespaces G[p_2] \; R \; yz \Lp \Trip{z}{\Ey}{L_2}}
			{G[p_1 \; p_2] \; R \; xyz \Lp \Trip{z}{\Suff{\Ex}{\Ey}}{L_1 \dplus L_2}}}   \mylabel{seq.1}
        \;\;\;
		{\frac{G[p_1] \; R \; xyz \Lp \Trip{yz}{\Ex}{L_1} \fivespaces G[p_2] \; R \; yz \Lp \Trip{\Lfail}{\Ey}{L_2}}
			{G[p_1 \; p_2] \; R \; xyz \Lp \Trip{\Lfail}{\Suff{\Ex}{\Ey}}{L_1 \dplus L_2}}}   \mylabel{seq.2}
		\\ \\
		& 
		 {\frac{G[p_1] \; R \; xyz \Lp \Trip{yz}{\Ex}{L_1} \fivespaces G[p_2] \; R \; yz \Lp \Trip{l}{z}{L_2} \fivespaces l \neq \Lfail}
			{G[p_1 \; p_2] \; R \; xyz \Lp \Trip{l}{z}{L_1 \dplus L_2}}}   \mylabel{seq.3}
		\;\;\;
		{\frac{G[p_1] \; R \; x \Lp \Trip{l}{\Ex}{L}}
			{G[p_1 \; p_2] \; R \; x \Lp \Trip{l}{\Ex}{L}}} \mylabel{seq.4}
		\\ \\
		& \textbf{Repetition} \;\;\;
		{\frac{G[p] \; R \; x \Lp \Trip{\Lfail}{\Ex}{L}}
			{G[p*] \; R \; x \Lp \Trip{x}{\Ex}{L}}} \mylabel{rep.1}
		\;\;\;
		{\frac{G[p] \; R \; xyz \Lp \Trip{yz}{\Ex}{L_1} \fivespaces G[p*] \; R \; yz \Lp \Trip{z}{\Ey}{L_2}}
			{G[p*] \; R \; xyz \Lp \Trip{z}{\Suff{\Ex}{\Ey}}{L_1 \dplus L_2}}} \mylabel{rep.2}
			\\ \\
		&
		\;\;\;
		{\frac{G[p] \; R \; xy \Lp \Trip{l}{y}{L} \fivespaces l \neq \Lfail}
			{G[p*] \; R \; xy \Lp \Trip{l}{y}{L}}} \mylabel{rep.3}
		\;\;\;
		{\frac{G[p] \; R \; xyz \Lp \Trip{yz}{\Ex}{L_1} \fivespaces G[p*] \; R \; yz \Lp \Trip{l}{z}{L_2}}
			{G[p*] \; R \; xyz \Lp \Trip{l}{\Suff{\Ex}{z}}{L_1 \dplus L_2}}} \mylabel{rep.4}    \\ \\
		& \textbf{Negative Predicate} \;\;\;
		{\frac{G[p] \; \{\} \; x \Lp \Trip{l}{\Ex}{[]}}
			{G[!p] \; R \; x \Lp \Trip{x}{\Nil}{[]}}} \mylabel{not.1}
		\;\;\;
		{\frac{G[p] \; \{\} \; xy \Lp \Trip{y}{\Ex}{[]}}
			{G[!p] \; R \; xy \Lp \Trip{\Lfail}{\Nil}{[]}}} \mylabel{not.2}
			\\ \\
		& \textbf{Ordered Choice} \;\;\;
		{\frac{\Matgk{p_1}{xy}{R} \Lp \Trip{y}{\Ex}{L}}
      {\Matgk{p_1 \;\slash\; p_2}{xy}{R} \Lp \Trip{y}{\Ex}{L}}} \mylabel{ord.1}
		\fivespaces
		{\frac{\Matgk{p_1}{xy}{R} \Lp \Trip{l}{y}{L} \fivespaces l \neq \Lfail}
      {\Matgk{p_1 \;\slash\; p_2}{xy}{R} \Lp \Trip{l}{y}{L}}} \mylabel{ord.2}
      \\ \\
		& {\frac{\Matgk{p_1}{xy}{R} \Lp \Trip{\Lfail}{\Ex}{L_1} \fivespaces \Matgk{p_2}{xy}{R} \Lp \Trip{\Lfail}{\Ey}{L_2}}
      {\Matgk{p_1 \;\slash\; p_2}{xy}{R} \Lp \Trip{\Lfail}{\Suff{\Ex}{\Ey}}{L_1 \dplus L_2}}} \mylabel{ord.3}
      \;\;\; {\frac{\Matgk{p_1}{xy}{R} \Lp \Trip{\Lfail}{\Ex}{L_1} \fivespaces \Matgk{p_2}{xy}{R} \Lp \Trip{y}{\Ey}{L_2}}
      {\Matgk{p_1 \;\slash\; p_2}{xy}{R} \Lp \Trip{y}{\Suff{\Ex}{\Ey}}{L_1 \dplus L_2}}} \mylabel{ord.4}
      \\ \\
     &	{\frac{\Matgk{p_1}{xy}{R} \Lp \Trip{\Lfail}{\Ex}{L_1} \fivespaces \Matgk{p_2}{xy}{R} \Lp \Trip{l}{y}{L_2} \fivespaces l \neq \Lfail}
      {\Matgk{p_1 \;\slash\; p_2}{xy}{R} \Lp \Trip{l}{y}{L_1 \dplus L_2}}} \mylabel{ord.5} 
		\\ \\
	& \textbf{Recovery} \fivespaces
	{\frac{l \notin Dom(R)}{\Matgk{\throw^{l}}{x}{R} \Lp \Trip{l}{x}{[]}}} \mylabel{throw.1}
	\;\;\;
	{\frac{\Matgk{R(l)}{xy}{R} \Lp \Trip{y}{\Ex}{L}}
      {\Matgk{\throw^{l}}{xy}{R} \Lp \Trip{y}{\Ex}{\Tup{l}{xy} :: L}}}  \mylabel{throw.2}
	\;\;\;
	{\frac{\Matgk{R(l_1)}{xy}{R} \Lp \Trip{l_2}{\Ex}{L}}
      {\Matgk{\throw^{l_1}}{xy}{R} \Lp \Trip{l_2}{\Ex}{\Tup{l_1}{xy} :: L}}} \mylabel{throw.3}
		\end{align*}
	}
	\caption{Semantics of PEGs with labels, recovery and farthest failure tracking}
	\label{fig:semrecovfar}
\end{figure*}

The labeled PEGs with farthest failure tracking we described in the previous
section make it easier to report the first syntax error found in a PEG, and
we will use them as the first step towards an error recovery mechanism.

Before giving the full formal definition of PEGs with error recovery,
let us return to the example program in Figure~\ref{fig:javaerror} and
its two syntax errors: a missing semicolon at the end of line 7,
and an extra semicolon at the end of line 8. The labeled PEG of
Figure~\ref{fig:javalabels} throws the label \texttt{semia} when it
finds the first error, and finishes parsing.

If every syntactic error is labeled, to recover from them we need to
do the following: first, catch the label right after it is thrown,
before the parser aborts, then log this error, 
possibly skip part of the input, and finally resume parsing.
In our example, for the first error we just need to log it and continue
as if the semicolon was found, and for the second error we need to log
the error, skip until finding the end of a block (taking care with
nested blocks on the way), and then resume.

To achieve this, we extend labeled PEGs with a list of recovered errors
and a map of labels to {\em recovery expressions}. These recovery
expressions are responsible for skipping tokens in the input
until finding a place where parsing can continue. 

Figure~\ref{fig:semrecovfar} presents the semantics of labeled PEGs
with error recovery as a set of inference rules for a $\Lp$
function. The notation $\Matgk{p}{xy}{R} \Lp \Trip{y}{v?}{L}$ represents a successful match of the parsing expression $p$ in the context of a PEG $G$ against the subject $xy$ with a map $R$
from labels to recovery expressions, 
consuming $x$ and leaving the suffix $y$. The term $\Ex$ is information for
tracking the location of the furthest failure, and denotes either a suffix $v$ of the {\em original} input or $\Nil$. $L$ is a list
of pairs of a label and a suffix of the original input, denoting
errors that were logged and recovered. For an unsuccessful
match the first element of the resulting triple is $l$, $f$,
or $fail$, denoting a label.

The auxiliary function \Suf\ that appears on
Figure~\ref{fig:semrecovfar} compares two possible
error positions, denoted by a suffix of the
input string, or \Nil\, if no failure has occurred,
and returns the furthest: any suffix of the input
is a further possible error position than \Nil\,
and a shorter suffix is a further possible
error position than a longer suffix.

Most of the rules are conservative extensions of the rules
for labeled PEGs~\cite{maidl2016peglabel,medeiros2016sac},
where the recovery map $R$ is simply passed along, and any
lists of recovered errors are concatenated. The exception
are the rules for the syntactic predicate and for throwing labels.

The syntactic predicate turns any failure label into a
success, using an empty recovered map to make sure that
errors are not recovered inside the predicate. Failure
tracking information is also thrown away. In essence,
any error that happens inside a syntactic predicate is
expected, and not considered a syntax error in the input.

The new rules {\bf throw.2} and {\bf throw.3} are where
error recovery happens. $R(l)$ denotes the recovery expression
associated with the label $l$.
When a label $l$ is thrown we check if $R$ has a recovery expression associated with it. If it does not ({\bf throw.1}) we just append the label and current
position to $L$ and propagate the error upwards so parsing aborts.
If a recovery expression is present we append the error and continue parsing with this expression ({\bf throw.2} and {\bf throw.3}).

The semantics of Figure 4
is conservative with regards to the semantics of both original
PEGs, as given by Ford~\cite{ford2004peg},
that is, a PEG that does not use the {\bf throw} operator and
does not have a recovery expression for {\tt fail} produces
the same result (failure or consuming a particular prefix of the input)
in both semantics. It is also conservative with regards to the
semantics of PEGs with
labels as given by Maidl et al.~\cite{maidl2013peglabel,maidl2016peglabel},
for expressions that do not have recovery expressions for any labels that
they throw. Proofs
of these propositions are straightforward inductions on the height
of the respective proof trees.

In our example from Figure~\ref{fig:javalabels}, we can recover
from a {\tt semia} error (as well as {\tt semip} and {\tt semid})
with a simple $\varepsilon$ recovery rule that matches the
empty string, wich will always succeed. This is similar to making
semicolons optional in the grammar, but recording that the
semicolon was not found instead of just ignoring the issue.

For the {\tt rcblk} error, our recovery needs an auxiliary rule in the grammar:
\begin{align*}
{\it SkipToRCUR} \leftarrow (!{\tt RCUR} \;({\tt LCUR} \; {\it SkipToRCUR} \; / \; .))* \; {\tt RCUR}
\end{align*}

This rule skips all tokens until finding and consuming a {\tt `\}'} ({\tt RCUR}) token, or reaching the end of input, taking care to correctly account for
nested blocks. One drawback of this recovery expression is that
it will make the parser ignore anything from the point of the
error to the closing brace, including any errors in that part of
the input. In the next section, we will discuss error recovery
strategies for PEGs, and how we can modify the grammar
to improve recovery of {\tt rcblk} errors.

\section{Error Recovery Strategies for PEGs}
\label{sec:strat}

A parser with a good recovery mechanism is
essential for use in an IDE, where we want an AST
that captures as much information as possible about
the program even in the presence of syntax errors
due to an unfinished program.

We can improve the error recovery quality of a
PEG parser by using the {\tt FIRST} and
{\tt FOLLOW} sets of parsing expressions when
throwing labels or recovering from an error.
A detailed discussion about {\tt FIRST} and
{\tt FOLLOW} sets in the context of PEGs can be found
in other papers~\cite{redz09,redz14,mascarenhas2014}.

In our grammar for a subset of Java, 
we can see that whenever rule
{\it Exp} is used it should be followed by
either a right parenthesis or a semicolon, so we could
define $(!({\tt RPAR} \;/\; {\tt SEMI}) \; .)*$ as a recovery expression, based on the {\tt FOLLOW} set of {\it Exp}.
Differently from the {\tt rcblk} recovery expression,
this one does not consume the synchronization symbols,
as they should be consumed by the following expression.

The recovery expression above could be automatically computed
from {\tt FOLLOW({\it Exp})} and associated with labels
{\tt condi}, {\tt condw}, {\tt edec}, {\tt rval}, {\tt eprint},
and {\tt parexp}. Another option is to compute a specific
{\tt FOLLOW} set for each use of {\it Exp}. For example,
the {\tt FOLLOW} set of the uses of {\it Exp} in {\it DecStmt}
and {\it AssignStmt} contains only {\tt SEMI}, while the {\tt FOLLOW} set of the uses of {\it Exp} in {\it IfStmt}, {\it WhileStmt}, {\it AtomExp}, and {\it PrintStmt} contains only {\tt RPAR}.

The use of the {\tt FOLLOW} set (probably enhanced by a
synchronization symbol such as {\tt `;'}) provides a default
error recovery strategy.
Let us apply this strategy for our annotated Java grammar and consider
that the Java program from Figure~\ref{fig:javaerror} has
an error on line 5, inside the condition of {\tt while} loop, as follows:
{\small
\begin{verbatim}
    5      while( < n) {
\end{verbatim}
}

Our default error recovery strategy will report this error
and resume parsing correctly at the following right parethensis.
In the resulting AST, the node for the {\tt while} loop will
have an empty condition, so we lose the node corresponding
to the use of the {\tt n} variable, and the information
that the condition was a {\tt <} expression.

Now let us consider this strategy for label {\tt rcblk}.
A {\it BlockStmt} can be followed by a {\it Statement},
or by {\tt RCURL}, so the default recovery expression associated
with {\tt rcblk} would synchronize with a token that
indicates the beginning of a {\it Statement}, 
an {\tt else} block, or with a {\tt `\}'}. 

Unfortunately,
this recovery strategy is not good for this error,
as our example program from Figure~\ref{fig:javaerror} shows.
The recovery expression for the {\tt rcblk} label will
consume the {\tt `;'} at the end of line 8, and then
stop at the beginning of the next statement on line 9.
But the parser just closed the {\it BlockStmt} of
the main function of the program, and now expects
another {\tt `\}'} to close the class body in {\it Prog}.
This will lead to a spurious error when the parser finds
the beginning of the print statement, so a custom {\it SkipToRCUR}
recovery expression is a better way to deal with {\tt rcblk}
errors.

While {\it SkipToRCUR} avoids spurious errors, it does have the
potential to skip a large portion of the input, leading to
a poor AST. We can improve
this by noticing that {\it Stmt} inside the repetition of {\tt BlockStmt}
is not allowed to fail unless the next token is {\tt RCUR}, 
so we can replace {\it Stmt} with $!{\tt RCUR} \; [{\it Stmt}]^{\tt stmtb}$. 
Now the second error in our program will make parsing fail with a {\tt stmtb} label. The recovery expression of this label
can synchronize with the beginning of the next
statement, or {\tt `\}'}. In our example, this will skip the
erroneous {\tt `;'} at the end of line 8 and then continue parsing the rest of the block. 

Finally, we have the full power of PEGs inside 
recovery expressions, and can use it for more elaborate
recovery strategies. Going back to the error in the
condition of a {\tt while} look earlier in this section,
we can, instead of blindly skipping tokens until finding
the closing {\tt `)'}, try to see if we have a partial
relational expression before giving up with the following
recovery expression for {\tt condw}:
{\small
\begin{align*}
    !!{\tt EQ} \; ({\tt EQ} \; [{\it RelExp}]^{\tt relexp})* \; / \;
     !!{\tt LT} \; ({\tt LT} \; [{\it AddExp}]^{\tt addexp})* \; / \;
     (!{\tt RPAR} \; .)*
\end{align*}}
The double negation is an {\em and} syntactic predicate,
and is a way of guarding an expression so it will only be
tried if its beginning matches the guard.

\section{Evaluation}
\label{sec:eval}

In this section, we evaluate our syntax error recovery approach
for PEGs using a complete parser for an existing programming language
in two different contexts, first in isolation and then by comparison
with a parser generated by a mature parser generator that uses predictive
parsing.

\subsection{Error recovery in a Lua parser}

It seems there is not a consensus about how to evaluate an error recovery strategy.
Ripley and Druseiks~\cite{ripley1978statistical} collected a set of syntactic
invalid Pascal programs that was used to evaluate some error recovery
strategies~\cite{pennello1978forward,burke1987practical,dain1994practical,corchuelo2002repair}.
However, as far as we know, this set of programs is not publicly available.

Another issue related to the evaluation of an error recovery strategy is how
to measure its quality. Pennelo and DeRemmer~\cite{pennello1978forward} proposed
a criteria based on the similarity of the program got after recovery with the
intended program (without syntax errors). This quality measure was used to evaluate several
strategies~\cite{corchuelo2002repair,degano1995comparison,dejonge2012natural},
although it is arguably subjective~\cite{dejonge2012natural}.
 
We will evaluate our strategy following Pennelo and DeRemmer's approach, however we will compare the AST got from an erroneous program after recovery with the AST of what would be the equivalent correct program, instead of comparing program texts. 

Based on this strategy, a recovery is {\it excellent}
when it gives us an AST equal to the intended one.
A {\it good} recovery gives us a reasonable AST, i.e.,
one that captures most information of the original program,
does not report spurious errors, and does not miss other errors.
A {\it poor} recovery, by its turn, produces an AST that
loses too much information, results in spurious errors, or
misses errors. Finally, a recovery is rated as {\it failed}
whenever it fails to produce an AST at all.

To evaluate our error recovery strategy, we built a PEG parser
for the Lua programming language~\cite{lua} using the LPegLabel 
tool, in which support for associating labels with recovery expressions has been added to its current version~\cite{lpeglabel}.
Our parser is based on the syntax defined in the Lua 5.3 reference
manual~\footnote{\url{https://www.lua.org/manual/5.3/}}, and builds
the AST associated with a given program.

We used 75 different labels to annotate the Lua grammar.
The process of annotating the Lua grammar with labels was done
manually, as well as the process of writing the recovery 
expressions for each label.

Initially, we defined a small set of default recovery expressions,
based on what would be good recovery tokens for the Lua grammar,
and we associated one of these expressions with each label of our
grammar. Then, while testing our recovery strategy we wrote some
custom recovery expressions in order to avoid spurious error messages
or to build a better AST. 

We wrote 180 syntactically invalid Lua programs to test our error
recovery mechanism. In a general way, each program should cause the
throwing of a specific label, to test whether the associated 
recovery expression recovers well. We usually wrote more than 
one erroneous Lua program to test each label.

Table~\ref{tab:eval1} shows for how many programs the recovery
strategy we implemented was considered {\it excellent},
{\it good}, {\it poor}, or {\it failed}. As we can see,
the use of labels plus the recovery operator enabled
us to implement a PEG parser for the Lua language with
a robust recovery mechanism. In our evaluation approach,
more than 90\% of the recovery done was considered
acceptable, i.e., it was rated at least {\it good}.

\begin{table}[t]
    \centering
    \begin{tabular}{|c|c|c|c|c|}
      \hline
     Excellent &            Good & Poor & Failed & Total \\ \hline
    100 ($\approx 56\%$) & 63 ($\approx 35\%$)   & 17 ($\approx 9\%$)   & 0    & 180 \\
          \hline
    \end{tabular}
    \caption{Evaluation of our Recovery Strategy Applied to a Lua Parser}
    \label{tab:eval1}
\vspace{-0.5cm}
\end{table}

Our parser was always able to build an AST, given that
no recovery expression raised an unrecoverable error,
or entered a loop. These
properties can be conservatively checked, as indicated by
Ford~\cite{ford2004peg}.


\subsection{Comparison with ANTLR}

ANTLR~\cite{antlrsite,parr2013antlr} is a popular tool
for generating top-down parsers. The repositoy of ANTLR
at GitHub contains the implementation of several parsers,
including a parser for Lua
5.3~\footnote{\url{https://github.com/antlr/grammars-v4/blob/master/lua/Lua.g4}}.

Unlike LPegLabel, ANTLR automatically generates 
from a grammar description a parser with error reporting and
recovery mechanisms , so the user does not need to annotate
the grammar. After an error, an ANTLR parser attempts single
token insertion and deletion to resynchronize. In case the 
remaining input can not be matched by any production of the 
current non-terminal, the parser consumes the input 
``\textit{until it finds a token that could reasonably follow
the current non-terminal}"~\cite{parr2014antlr}.

The available Lua parser based on ANTLR does not build an AST,
so we could not evaluate its recovery quality by comparing the
AST built by it with the AST built by the Lua parser generated
by LPegLabel. In order to compare the error reporting and recovery 
strategies of both parsers, we counted the number of error messages
generated by them for the 180 syntactic invalid Lua programs that
we used to test our Lua parser based on labeled PEGs with recovery 
expressions.

Table~\ref{tab:eval2} shows that for most programs both parsers
report the same number of errors. When the parsers report a
different amount of errors, usually the ANTLR parser reports
more errors than the LPegLabel one.

\begin{table}
\centering
\begin{tabular}{|c|c|}
  \hline
    Parser & \# of files \\ \hline
    ANTLR parser reported more errors &  56 ($\approx 31\%$)\\ \hline
    PEG parser reported more errors & 14 ($\approx 8\%$)\\ \hline
    Parsers Reported the same number of errors & 110  ($\approx 61\%$) \\
      \hline
\end{tabular}
\caption{Comparison of ANTLR-based and PEG-based Lua parsers with error recovery}
    \label{tab:eval2}
\vspace{-0.5cm}
\end{table}

There are two possibilities that could explain why the LPegLabel parser
gives less error messages: 
\begin{enumerate}
    \item After an error, the LPegLabel parser usually discards more input
    than the ANTLR one, possibly skiping other errors.
    \item After an error, the LPegLabel usually syncronizes well, avoiding
    spurious errors.
\end{enumerate}

Previously, in Table~\ref{tab:eval1}, we have seen that the Lua parser
based on LPegLabel often builds a good AST, so we can say that option 1
is not a good explanation. We can state with confidence that, for this
set of syntactic invalid programs, the PEG-based parser reports less errors
than the ANTLR one because the latter is producing more spurious errors.

Moreover, since the sychronization strategy of the PEG-based parser
is manually designed by the user, it is expected that it synchronizes
better after an error. Nevertheless, this is still evidence that
our approach based on labels and recovery expressions is effective,
as having the full power of parsing expression grammars available
when writing recovery expressions makes it easy to tailor the recovery
strategies for each kind of error.

For example, let us consider the following Lua program where the
user did not type the condition between an \texttt{if} and the corresponding
{\tt then}:
\begin{verbatim}
    if then print("that") end
\end{verbatim}

The ANTLR parser gives us the following error messages:
\begin{verbatim}
line 1:7 extraneous input 'then' expecting {'function', 
  'nil', 'false', 'true', [15 more tokens] }
line 1:26 missing 'then' at 'end'
\end{verbatim}

The first message correctly indicates the error position, but does not help
much to fix the error, as the programmer has to infer that the fifteen
tokens that the error message lists are tokens that begin expressions.
The second error message is spurious, a side effect of the parser skipping {\tt then}
and using {\tt print("that")} as the condition.

Our PEG-based parser reports a single error with error
message 
``{\tt syntax error, expected a condition after 'if'}" at
column 4, which seems more helpful to the programmer, and
correctly parses the rest of the {\tt if} statement.

We also compared the performance of the Lua parser generated by ANTLR
with the performance of our PEG-based parser. We used the following
tools in our comparison:
\begin{itemize}
    \item ANTLR 4.6 and 4.7, with Java OpenJDK 9
    \item LPegLabel 1.4, with Lua 5.3 interpreter
\end{itemize}

The test machine was an Intel i7-4790 CPU with 16G RAM, running
Ubuntu 16.04 LTS desktop.

We made two tests. In the first test, we created an invalid Lua program
\texttt{broke.lua} that was formed by concatenating almost all
the 180 erroneous programs that we have used before.
This file has around 550 lines,
and both parsers report more than 200 syntax errors while parsing it.
This file was used to measure the performance of parsers in a
syntactic invalid program. The Lua parser generated by ANTLR 4.7
crashed when parsing this file, so for this comparison we used a Lua
parser generated by ANTLR 4.6.

We also used both parsers to parse the test files from the Lua
5.3.4 distribution~\footnote{\url{https://www.lua.org/tests/lua-5.3.4-tests.tar.gz}}. The test comprises 28 syntactic valid Lua programs, which
toghether have more than 12k source lines. We needed to change the first
line of test file \texttt{main.lua}, because the Lua parser generated
by ANTLR could not recognize it.

We ran both parsers 20 times and collected the time reported
by {\tt System.nanoTime} for ANTLR and by {\tt os.time} for Lua.
For ANTLR, we measured the time by using {\tt @init} and
{\tt @after} actions in the start rule of the grammar.
In the case of LPegLabel, we measured the time before and after
calling the main function of the parser.
Tables~\ref{tab:broke} and~\ref{tab:luatest} show our results.
We can see that the PEG-based parser was significantly
faster (by approximately a factor of six) than the ANTLR parser in
both tests.     

\begin{table}[t]
    {\begin{tabularx}{0.45\textwidth}{|c| *{2}{>{\centering\arraybackslash}X|c|}}
      \hline
      & Avg & Median & SD  \\
      \hline
      LPegLabel & 14 & 15 & 0.4 \\ \hline
      ANTLR & 89 & 85 & 11 \\
      \hline
      \end{tabularx}}
    {\caption{Time (in ms) to parse broke.lua}
      \label{tab:broke}}
\vspace{-0.5cm}
\end{table}

\begin{table}[t]
    {\begin{tabularx}{0.45\textwidth}{|c| *{2}{>{\centering\arraybackslash}X|c|}}
      \hline
      & Avg & Median & SD \\
      \hline
      LPegLabel & 94 & 94 & 1 \\ \hline
      ANTLR & 647 & 624 & 76 \\
      \hline
      \end{tabularx}}
    {\caption{Time (in ms) to parse Lua 5.3 tests}
      \label{tab:luatest}}
\vspace{-0.5cm}
\end{table}


\section{Related Work} 
\label{sec:rel}

In this section, we discuss other error recovery approaches
used by top-down parsers with backtracking, focusing on PEGs.
Error handling in top-down parsing based on CFGs is a well-studied subject. 
Grune and Jacobs~\cite{grune2010ptp} presents an overview of several
error handling techniques used in this context.

Swierstra and Duponcheel~\cite{swierstra1996dec} shows an
implementation of parser combinators for error recovery,
but is restricted to LL(1) grammars. The recovery strategy
is based on a {\em noskip} set, computed by taking the
{\it FIRST} set of every symbol in the tails of the pending
rules in the parser stack. Associated with each token in this set
is a sequence of symbols (including non-terminals) that would
have to be inserted to reach that point in the parse, taken from
the tails of the pending rules. Tokens are then skipped until
reaching a token in this set, and the parser then takes actions
as if it found the sequence of inserted symbols for this token.

Our approach cannot simulate this recovery strategy, as it relies
on the path that the parser dynamically took to reach the point of
the error, while our recovery expressions are statically determined
from the label. But while their strategy is more resistant to
the introduction of spurious errors than just using the {\tt FOLLOW}
set it still can introduce those.

A common way to implement error recovery in PEG parsers
is to add an alternative to a failing expression,
where this new alternative works as a fallback. Semantic actions
are used for logging the error.
This strategy is mentioned in the manual of Mouse~\cite{redzmouse}
and also by users of LPeg~\footnote{See 
\url{http://lua-users.org/lists/lua-l/2008-09/msg00424.html}}.
These fallback expressions with semantic actions for error logging
are similar to our labels and recovery expressions, but in an ad-hoc,
implementation-specific way.

Several PEG implementations such as
Parboiled~\footnote{\url{https://github.com/sirthias/parboiled/wiki}},
Tatsu~\footnote{\url{https://tatsu.readthedocs.io}},
and PEGTL~\footnote{\url{https://github.com/taocpp/PEGTL}} provide 
features that facilitate error recovery.

Parboiled uses an error recovery strategy based on ANTLR's one.
When the input is not valid, Parboiled parses it again
to determine the error location, then it does another
parse and tries to recover from the error by including,
removing, or replacing one input character from the
error position. In case this strategy does not work,
Parboiled parses the input once more and automatically
chooses a resynchronization rule based on the sequence
of parsing rules used until the error position.

Similar to ANTLR, the strategy used by Parboiled 
is fully automated, and requires neither manual intervention
nor annotations in the grammar. Our approach
currently requires grammar annotations to be fully effective,
but the work of inserting this annotations can be automated
in several cases. On the other hand, we do not require
parsing the input multiple times, which leads to better performance.

Tatsu uses the fallback alternative technique for error
recovery, with the addition of a {\it skip expression},
which is a syntactic sugar for defining a pattern that
consumes the input until the skip expression succeeds. 

PEGTL allows to define for each rule $R$ a set of
terminator tokens $T$, so when the matching of $R$
fails, the input is consumed until a token $t \in T$
is matched. This is also similar to our approach for recovery
expressions, but with coarser granularity, and lesser control
on what can be done after an error.

Rüfenacht~\cite{michael2016error} proposes a local
error handling strategy for PEGs. This strategy uses
the farthest failure position and a record of the parser
state to identify an error. Based on the information
about an error, an appropriate recovey set is used.
This set is formed by parsing expressions that match
the input at or after the error location, and it is used
to determine how to repair the input. 

The approach proposed by Rüfenacht is also similar to the use
of a recovery expression after an error, but more limited
in the kind of recovery that it can do. When testing his approach in
the context of a JSON grammar, Rüfenacht noticed long running
test cases and mentions the need to improve memory use and
other performance issues.

\section{Conclusions} 
\label{sec:conc}

We have presented a conservative extension of PEGs 
that is well-suited for implementing parsers with a robust
mechanism for recovering from syntax errors in the input.
Our extension is
based on the use of labels to signal syntax errors, and
differentiates them from regular failures, together
with the use of recovery expressions associated with
those labels.

When signaling an error with a label that has an associated recovery 
expression, the parser logs the label and the error position, then
proceeds with the recovery expression. This recovery
expression is a regular parsing expression, with access to all
the parsing rules that the grammar provides.

We tested our recovery mechanism by implementing it on the
current version of LPegLabel,
an existing parser generator for labeled PEGs, and used this
tool to create a parser with error recovery for the Lua programming 
language. We tested this parser on a suite of 180 programs with
syntax errors to assess how close the syntax trees produced
by our parser are to the trees we get from manually fixing the
syntax errors present in the programs. Our parser gives at least
good results for 91\% of our test programs, and excellent results
for 56\% of them.

We also compared our parser with a Lua parser with automatic error
recovery generated by ANTLR, a popular parser generator tool.
The comparison shows that our PEG-based parser has better error recovery,
error messages, and better performance than the ANTLR-generated one.

Labeled PEGs with recovery expressions give the grammar writer
great control over the error recovery strategy, at the cost of an
annotation burden that we judge to not be too onerous. Nevertheless,
we want to study ways of automating label insertion, as well as
generation of good recovery expressions.

Finally, our evaluation did not try to take into account errors
that are more frequent while writing Lua programs from scratch
in the context of an IDE or text editor. Such a study can also
be explored in future work.

\bibliographystyle{ACM-Reference-Format}
\bibliography{recovery}

\end{document}